\documentstyle[12pt]{article}
\renewcommand{\theequation}{\arabic{section}.\arabic{equation}}

\newcommand{\be}{\begin{equation}}
\newcommand{\ee}{\end{equation}}
\newcommand{\ba}{\begin{array}}
\newcommand{\ea}{\end{array}}
\newcommand{\bc}{\begin{center}}
\newcommand{\ec}{\end{center}}
\newcommand{\disregard}[1]{{}}
\newcommand{\ti}{\tilde}
\newcommand{\al}{\alpha}
\newcommand{\am}{\langle\alpha\rangle}
\newcommand{\ord}{\Phi}
\newcommand{\nc}{n_c}
\newcommand{\nt}{\ti{n}}
\newcommand{\mt}{\ti{m}}

\newcommand{\ds}{\displaystyle}
\newcommand{\dt}{\hbox{d}t}
\newcommand{\demi}{{\ds 1\over\ds 2}}
\newcommand{\quart}{{\ds 1\over\ds 4}}
\newcommand{\ki}{\xi}
\newcommand{\ep}{\epsilon_k}
\newcommand{\epk}{\epsilon_k^{'}}

\newcommand{\apb}[1]{Ann. of Phys. (N.Y.) {\bf #1} }
\newcommand{\npb}[1]{Nucl. Phys. {\bf #1}}
\title{{\it BEC} From a Time-Dependent Variational Point of View}

\author{
Mohamed Benarous
\\
{\it Laboratory for Theoretical Physics and Material Physics}
\\
{\it Faculty of Sciences and Engineering Sciences}
\\
{\it Hassiba Benbouali University of Chlef}
\\
{\it B.P. 151, 02000 Chlef, Algeria.}
}

\date{\today}
\begin{document}
\maketitle
\begin{abstract}
We use the time-dependent variational principle of Balian and V\'en\'eroni to
derive a set of equations governing the dynamics of a trapped Bose gas at finite 
temperature.
We show that this dynamics generalizes the Gross-Pitaevskii equations in that
it introduces a consistent dynamical coupling between the evolution of the condensate density, 
the thermal cloud and the ''anomalous'' density.

\end{abstract}


\newpage

\setcounter{section}{0}

\setcounter{equation}{0}
\bc{\section{Introduction}} \ec

Bose-Einstein condensation (BEC) phenomenon has a long story since its
discovery.
Such a fascinating behavior of matter has intrigued many researchers\cite{BE}.
Static as well as dynamic properties of BEC are now intensively studied,
both experimentally and theoretically, in order to apprehend the way the
BEC forms, develops and vanishes.

Various theoretical techniques have been used successfully,
predicting correctly a great number of experimental results. Among these
techniques, some are most widely used, such as the Bogoliubov\cite{BG},
the Beliaev\cite{BL,GR} and the Hartree-Fock-Bogoliubov\cite{HF,KT97,ST} approximations.
Some other methods, such as the Monte-Carlo simulations\cite{KR} are rising
interest since they tend to solve the exact quantum many-body problem.

Although being extremely precise in the static situations, the approximations
mentioned above all adopt ad-hoc assumptions about the various quantities
such as the order parameter $\ord$ (or the condensate density $\nc$),
the non-condensed density or thermal cloud $\nt$ or the anomalous density $\mt$. 
These assumptions lead inevitably to the fact that the approximations are no longer
totally consistent in a sense which will become clearer soon.

In this paper, we rely on a different approach, based on the time-dependent
variational principle of Balian and V\'en\'eroni (BV) \cite{BV,BF99}.
Not only does this method retain the essential features of the physics
behind the previous approximations, but it also allows one to bypass some
(if not all) of the ad-hoc assumptions. Indeed, being variational, the
formalism generates a consistent dynamics for the quantities $\ord$,
$\nc$, $\nt$ and $\mt$ by choosing a certain class of trial spaces.

This well-known advantage of this (and any) variational principle faces
however a major difficulty related to the choice of the trial spaces.
A (difficult) compromise must be found between a correct description
of the physics on one hand, and the tractability of the calculations
on the other. In this order of ideas, the BV variational principle requires
to specify two objects: a density-like operator and a ''measured'' observable.

Our choice for the variational spaces consists of a gaussian time-dependent
density-like operator and a single-particle operator for the observable.
This last quantity turns out to be of no interest in our particular case
but it may play a major role in other situations especially when one intends
to calculate correlation functions\cite{BV93,BM98}.

The machinery we are discussing has in fact already been used
elsewhere\cite{BF99}, where we have recognized that the variational
dynamical equations derived there are a generalization of the
Gross-Pitaevskii equations\cite{GP}, that takes into account the coupling
between the order parameter and the various densities. We called this
approach the ''Time-Dependent Hartree-Fock-Bogoliubov'' (TDHFB) approximation. 
The point is that the usual image of a collection of condensed atoms in a bath 
of thermal particles is not totally true, since the bath has its own dynamics 
which is sensible to the condensate dynamics.

The purpose of the present paper is to go into further details in this
variational approximation so as to make a closer connection with other
well-known methods (such as the ones quoted above) used in the study of
Bose-Einstein condensates. Among the problems that we intend to study,
we can cite in particular the analysis of the static properties of the
condensate (compared to what is known) as well as the small oscillations
around the static solutions, when the effects of the coupling with the
thermal cloud and the anomalous density are taken into account.

The important paper by A. K. Kerman and P. Tommasini \cite{KT} is closely related to ours. 
It uses however the Dirac variational principle which constrains the calculations to $T=0$
(even if the authors give at the end of the paper a finite temperature prescription.) 
The authors also limit themselves to the uniform (that is non trapped) case. Therefore,
according to what has been said above, we may consider our paper as a twofold generalization 
of \cite{KT}.

The full TDHFB equations also deserve to be solved in order
to study the large amplitude motion of the condensate and the thermal cloud.
Despite its importance, we will postpone this study to a future publication 
which is in progress.

The paper is organized as follows. In section 2, we recall the major steps
used in \cite{BF99,BM98} to derive the TDHFB equations and to write them 
down in the case of the BEC problem. Section 3 is devoted to the study of 
the static solutions, where we recover the results of \cite{GR} and generalize 
them to finite temperature. In section 4, we analyze the excitations of the 
condensate by using the RPA technique. Finally, we present some concluding 
remarks and perspectives in order to deal with more complicated situations.

\setcounter{equation}{0}
\bc{\section{The TDHFB Equations}}\ec

The General TDHFB equations were derived in \cite{BF99} using the BV
variational principle. For technical reasons, they were written in
terms of the creation and annihilation operators $a^{+}$ and $a$.
They may however easily be translated in a more appropriate notation for
the BEC problem using the boson field operator $\hat{\Psi}({\bf r})$ in the
Schr\"o\-dinger picture, in exactly the same way as our previous
work\cite{BM98}.
Let us first recall some of our notations. We introduced the gaussian density
operator ${\cal D}(t)$ (with variational parameters ${\cal N}(t)$,
$\lambda (t)$ and $S (t)$):
\be\label{eq1}
{\cal D} (t) = {\cal N}(t) \exp{(\lambda (t)\tau\alpha )}
\exp{(\demi\alpha\tau S (t)\alpha )},
\ee
where $\alpha = (a^{+},a)$ and $\tau$ is a symplectic constant matrix.
Then, we defined the one-boson expectation value $\langle\alpha\rangle$
and the single-particle density matrix $\rho$ by:
\be\label{eq2}
\langle\alpha\rangle =
\left(\ba{c}
\langle a\rangle \\
\langle a^{+}\rangle \ea\right) \quad,\quad
1+2\rho = \pmatrix{
\langle \bar{a}\bar{a}^{+}+\bar{a}^{+}\bar{a}\rangle
&
-2\langle\bar{a}\bar{a}\rangle
\cr
2\langle\bar{a}^{+}\bar{a}^{+}\rangle
&
-\langle \bar{a}\bar{a}^{+}+\bar{a}^{+}\bar{a}\rangle
\cr},
\ee
which are directly related to $\lambda (t)$ and $S (t)$ respectively
(see Eq. A.6 of ref.\cite{BM98}). Operators such as $\bar a$  are simply
the centered operators $a-\langle a\rangle$. The expectation values are
to be taken with respect to the density operator (\ref{eq1}). 


Introducing (\ref{eq1}) in the BV variational action-like leads, beside
the conservation of the partition function ${\cal Z}=\hbox{Tr}\,{\cal D}(t)$,
to what we called the TDHFB equations:
\be\label{eq3}
\ba{rl}
i\hbar {\ds\hbox{d}\langle\alpha\rangle\over\ds\dt} & = \tau
{\ds\partial\langle H\rangle\over\ds\partial\langle\alpha\rangle }
,\\
i\hbar {\ds\hbox{d}\rho\over\ds\dt} & = -2 \left[\rho ,
{\ds\partial\langle H\rangle\over\ds\partial\rho}\right]
\ea ,
\ee
in which $\langle H\rangle$ is the energy. Some interesting properties are
discussed in (\cite{BM98,BF99}).

In order to make connection with the BEC phenomenon, we introduce first
the Hamiltonian for trapped bosons\cite{GR}:
\be\label{eq4}
H =\int_{\bf r} a^{+} ({\bf r})\left[-{\ds\hbar^2\over\ds 2m}\Delta +
V_{\rm{ext}}({\bf r})-\mu\right]a({\bf r})+{\ds g\over\ds 2}\int_{\bf r}
a^{+}({\bf r})a^{+}({\bf r})a({\bf r})a({\bf r}),
\ee
where $V_{\rm{ext}}({\bf r})$ is the trapping potential, $\mu$ is the
chemical potential and $g$ is the coupling constant. The energy
${\cal E}=\langle H\rangle$ is easily computed yielding:
\be\label{eq5}
\ba{rl}
{\cal E} = {\ds\int_{\bf r}}
[
& -{\ds\hbar^2\over\ds 2m}\ord^{*}\Delta\ord -{\ds\hbar^2\over\ds
2m}\ord\Delta\ord^{*}
+ (V_{\rm ext}-\mu +2g\nt)|\ord |^2 +{\ds g\over\ds 2}|\ord |^4
-{\ds\hbar^2\over\ds 2m}\nt\Delta
\\
&+(V_{\rm ext}-\mu )\nt + g \nt^2 + {\ds g\over\ds 2}(|\mt |^2 + \mt^{*}\ord^2 + \mt {\ord^{*}}^2)
]
\ea \\,
\ee
where the condensate density $\nc = |\ord |^2$, the non-condensed density
$\nt$ and the anomalous density $\mt$ are identified respectively with
$|\langle a\rangle|^2$, $\langle\bar {a}^{+}\bar {a}\rangle$ and
$\langle\bar {a}\bar {a}\rangle$. This is simply because we identify
in our formalism the boson field operator ${\hat\Psi}$ with the destruction
operator $a$ and its fluctuation $\tilde\Psi$ with $\bar a$.

With these identifications, the Eqs.(\ref{eq3}) now take the form
\be\label{eq6}
\ba{rl}
i\hbar \dot{\ord} & = {\ds\partial {\cal E}\over\ds\partial\ord^{*} }
,\\
i\hbar \dot{\nt}  & = 2 \left(\mt^{*}{\ds\partial {\cal E}\over\ds\partial
\mt^{*} }
- \mt {\ds\partial {\cal E}\over\ds\partial\mt }\right)
,\\
i\hbar \dot{\mt} & = 4(\nt +\demi ){\ds\partial {\cal E}\over\ds\partial\mt^{*} }
+4{\ds\partial {\cal E}\over\ds\partial\nt}\mt ,
\ea
\ee
which constitutes a closed self-consistent system. The coupling between
the order parameter $\ord$, the non-condensed density and the anomalous density
occurs via the derivatives of $\cal E$ which still contain $\nt$ and $\mt$.

Beside the conservation of the energy, the equations (\ref{eq6}) exhibit the
unitary evolution of the density matrix (already visible in (\ref{eq3})) by
means of the conservation of the ''Heisenberg parameter'' $I$ defined as
$\rho (\rho +1)={\ds I-1\over\ds 4}$, or equivalently
\be\label{eq6a}
I=(2\nt +1)^2-4|\mt |^2.
\ee
We recall the reader that $I=\coth^2 (\hbar\omega /kT)$ for a thermal distribution.

The expression (\ref{eq5}) for the energy allows us to write down the
Eqs.(\ref{eq6}) more explicitly. They indeed read
\be\label{eq7}
\ba{rl}
i\hbar \dot{\ord} & = \left(
-{\ds\hbar^2\over\ds 2m}\Delta + V_{\rm{ext}}-\mu +g\nc+2g\nt\right)\ord
+ g\mt\ord^{*}
,\\
i\hbar \dot{\nt}  & = g\left(\mt^{*}\ord^2-\mt {\ord^{*}}^2\right)
,\\
i\hbar \dot{\mt} & = g (2\nt +1)\ord^2 + 4\left(
-{\ds\hbar^2\over\ds 2m}\Delta + V_{\rm{ext}}-\mu+2g n+{\ds g\over\ds 4}(2\nt +1)
\right)\mt
.
\ea 
\ee
The consistency of our derivation mentioned in the introduction is now clear.
Indeed, we obtain in Eqs.(\ref{eq7}) a self-consistent dynamics of
the order parameter, the thermal cloud and the anomalous density.
The equation governing the evolution of $\ord$ has been obtained elsewhere
\cite{GR,PO,J96,M97,Gi98} as an extension of the Gross-Pitaevskii equation,
but to our knowledge, the two last equations in (\ref{eq7}), governing
the evolution of $\nt$ and $\mt$, were never written down before at finite 
temperature.
The exception is ref.\cite{KT}, where a zero temperature (that is
$I=1$) and uniform ($V_{\rm{ext}}=0$) version was derived. Indeed, taking
these limits in (\ref{eq6a}) and (\ref{eq7}) provides the equations obtained
in ref.\cite{KT}. Therefore, our equations are more general.
They describe not only a self-consistent dynamics of the
non-condensed and anomalous densities but also a feedback effect on the order
parameter and therefore on the condensate density. The coupling is however 
intimately related to the two-body interactions and completely disappears for 
noninteracting systems, justifying therefore the Popov and the Beliaev approximations
for weakly interacting atomic gazes.

It is worth noticing that this dynamics is also number conserving since the total 
density $n=\nc +\nt$ is preserved during the evolution.

As a final remark, we may note that the information on the temperature is
encoded in the equation (\ref{eq6a}) which is a property of the density
operator (\ref{eq1}). Indeed, if $T$ is specified, then the Heisenberg
parameter $I$ (which we recall is a conserved quantity) is calculable and
this in turn allows the computation of $\nt$ (respectively $\mt$) once $\mt$
(respectively $\nt$) is known. The last two equations in (\ref{eq7}) are
therefore not totally independent.


\setcounter{equation}{0}
\bc{\section{The Static Solutions}}\ec

The static properties of the BEC at $T=0$ are well known\cite{GR}.
Indeed, at zero temperature, all the atoms are condensed. Therefore,
$\nt0=0$ and $\mt0=0$ and $\nc0$ equals the total density of the gas $n$.
The last two equations in (\ref{eq7}) become meaningless, and the first one
gives:
\be\label{eq9}
\left(-{\ds\hbar^2\over\ds 2m}\Delta + V_{\rm{ext}}-\mu +g\nc0
\right)\ord0 = 0.
\ee
This provides the density profile in the Thomas-Fermi (TF) approximation\cite{SV}:
\be\label{eq10}
\nc0 (r) = {\ds 1\over\ds g}\left(\mu -V_{\rm{ext}}(r)\right).
\ee
Upon defining the ''size'' of the fundamental state
$a_{H0}=(\hbar /m\omega0)^{1/2}$ and the s-wave scattering length
$a=mg/4\pi\hbar^2$, we obtain for a spherical potential
$V_{\rm{ext}}(r)=\demi m\omega0^2 r^2$, the condensate radius
$R_{\rm{TF}}$ and the chemical potential $\mu$ for a gas of $N0$ bosons as
\be\label{eq11}
\ba{rl}
R_{\rm{TF}} & = a_{H0}\left(15 N0 {\ds a\over\ds a_{H0}}\right)^{1/5}
,\\
\mu & = \demi\hbar\omega0 \left(15 N0 {\ds a\over\ds a_{H0}}\right)^{2/5}
,
\ea 
\ee
which are well-known expressions in the literature\cite{GR,ST}.

At $T>T_{\rm{BEC}}$, there is no condensate so that $\nc0 =0$ and $\nt0 =n$.
The static TDHFB equations reduce to a single equation which becomes in the TF
approximation
\be\label{eq12}
\left(V_{\rm{ext}}(r)-\mu +10gn +g\right)\mt0 = 0,
\ee
and leads to either a uniform density profile (for $\mt0 =0$), which seems
unreasonable in the presence of a confining field, or to a non-uniform
density of the form
\be\label{eq13}
n = {\ds 1\over\ds 10}\left({\ds\mu -V_{\rm{ext}}(r)\over\ds g}-1\right),
\ee
giving rise to the anomalous density
\be\label{eq14}
|\mt0 |^2 = \left(n -{\ds\sqrt{I}-1\over\ds 2}\right)
\left(n +{\ds\sqrt{I}+1\over\ds 2}\right)
.
\ee
The maximum spreading of the trapped gas $R$ as well as the chemical potential $\mu$ 
can be computed from (\ref{eq13}). We obtain the results
\be\label{eq15}
\ba{rl}
R & = a_{H0}\left(150 N0 {\ds a\over\ds a_{H0}}\right)^{1/5}
,\\
\mu & =g+ \demi\hbar\omega0 \left(150 N0 {\ds a\over\ds a_{H0}}\right)^{2/5}
,
\ea 
\ee
where we can notice the similarities with the totally condensed phase. In the two cases,
the spreading of the condensate (or of the global system for $T>T_{\rm{BEC}}$) depends 
essentially on the balance between the self-interactions and the confining potential.
Furthermore, we note on (\ref{eq14}) that at a maximum distance $R_{\rm{max}}$ 
from the center of the trap
\be\label{eq16}
R_{\rm{max}} = R\sqrt{1-{\ds 10g\over\ds m\omega0^2 R^2}(\sqrt{I}-1)} ,
\ee
the anomalous density vanishes. 

When $0<T<T_{\rm{BEC}}$, we have of course $\nc0\neq 0$ and $\nt0\neq 0$. In the TF 
approximation, we see that the value $\mt0 =0$ can no longer be retained. Therefore, 
the anomalous density, although (maybe) small, is an essential ingredient in the resolution 
of the static equations.

It is important to notice at this stage that the TF approximation is a somewhat hazardous 
hypothesis for the thermal cloud\cite{schuck}. Indeed, the traditional image of a condensate 
surrounded by a smooth thermal cloud is a rather simplified picture. Fortunately, we see on 
Eqs.(\ref{eq7}) that $\nt$ can be eliminated in favor of the ''relevant '' variables $\nc$ 
and $\mt$. We will henceforth mean by TF approximation the neglect of the kinetic terms 
in the equations of $\nc$ and $\mt$. 

Let us set $\ki ={\ds 1\over\ds g}\left(V_{\rm{ext}}(r)-\mu\right)$ 
and introduce the parametrization $\mt0 =|\mt0|\exp{(i\alpha)}$ and 
$\ord0 =\sqrt{\nc0}\exp{(i\phi)}$. We then obtain the implicit solutions:
\be\label{eq18}
\ba{rl}
|\mt0| & = \quart\left[{\ds I\over\ds 1-2q}-(1-2q)\right]
\\
\nt0  & = -\demi +\quart\left[{\ds I\over\ds 1-2q}+(1-2q)\right]
\\
\nc0 & = 1-\ki-\quart\left[{\ds I\over\ds 1-2q}+3(1-2q)\right]
,
\ea 
\ee
with $q=\ki +n$. This is obtained by setting $\alpha = 2\phi +\pi$ which is 
compatible with the Hugenholtz-Pines theorem \cite{HP} expressed in our 
context by the identity $|\mt0|=\ki +n+\nt0$.

The equations (\ref{eq18}), together with the third static equation in (\ref{eq7}) may be 
combined to yield a quartic equation for $q$ alone which can then be solved numerically to 
provide temperature and position-dependent density profiles. An important preliminary result 
is that the anomalous density is always very small compared to $\nc0$ or $\nt0$ whatever 
the conditions are, therefore, justifying a posteriori, the TF approximation used above. 
We shall discuss this and other numerical results in a separate work\cite{BC}. 

Nevertheless, one can gain further insights into the static properties at $0<T<T_{\rm{BEC}}$ by
choosing the parametrization
\be\label{eqst1}
\ba{rl}
1+2\nt0 & = \sqrt{I}\cosh{\sigma}
\\
2|\mt0| & = \sqrt{I}\sinh{\sigma}
,
\ea 
\ee
which automatically satisfies (\ref{eq6a}). Indeed, since we know that the $T=0$ case is given 
by $\sigma=0$, we may approximately solve the static problem to first order in $\sigma$. This is
equivalent to a low temperature expansion (but far from the transition). 
After some algebra, we obtain
\be\label{eqst2}
\ba{rl}
\nc0   & = -\ki-(\sqrt{I}-1)+\sqrt{I}\,\eta
\\
\nt0   & = {\ds \sqrt{I}-1 \over\ds 2}+\sqrt{I}\,\eta^2
\\
|\mt0| & = -\sqrt{I}\,\eta (1+2\eta)
,
\ea 
\ee
$\eta$ being a small expansion parameter. What we observe on (\ref{eqst2}) is that it is a 
natural extension to finite temperature of an expression like (\ref{eq10}). To lowest order 
($\eta=0$), the result is a temperature-dependent shift of the condensate density with 
respect to the $T=0$ case. 

Let us see what happens to the condensate radius and the chemical potential which were given 
by (\ref{eq11}). We define the condensate radius in the TF approximation by the point where 
$\nc0$ vanishes. This gives
\be\label{eqst3}
V_{\rm ext}(R_{\rm TF})=\mu-(\sqrt{I}-1)g.
\ee
The number of condensed atoms $N_c$, which is a measurable quantity, writes
\be\label{eqst4}
N_c = 4\pi\int0^{R_{\rm TF}} \nc0 (r) r^2 dr.
\ee
After integrating and using (\ref{eqst3}), we obtain the following remarkable
expressions:
\be\label{eqst5}
\ba{rl}
R_{\rm{TF}} & = a_{H0}\left(15 N_c {\ds a\over\ds a_{H0}}\right)^{1/5}
,\\
\mu & = g(\sqrt{I}-1)+\demi\hbar\omega0
\left(15 N_c {\ds a\over\ds a_{H0}}\right)^{2/5}
.
\ea
\ee
The condensate radius is thus given by the same formula as its zero
temperature counterpart, the sole difference lying in the appearance of
$N_c$ instead of the total number $N0$. For the chemical potential,
we observe, as for $\nc0$, a temperature-dependent shift with respect
to the $T=0$ case, but here also, it is the number of condensed atoms
which is involved and not the total number of atoms.

\disregard{

The Eqs.(\ref{eq18}) . 
This reads (in terms of $X=1-2q$)
\be\label{eq19}
2X^4 -3(1-\ki)X^3+IX^2-(1-\ki)IX+I^2=0,
\ee
and can be solved to yield a temperature-dependent density profile $n(r,T)$. 
In spite of the fact that the numerical analysis is still under work, some 
preliminary results can be presented here.
Nevertheless, one may go further in the analytic treatment by focusing as above on the 
condensate radius $R_{\rm{TF}}$. This quantity is determined by the condition 
$\nc0 (R_{\rm{TF}})=0$. Upon defining the transition temperature $T_{\rm{BEC}}$ as 
the point where the condensate collapses (that is, $R_{\rm{TF}}=0$), we find using the 
previous spherical potential
\be\label{eq20}
R_{\rm{TF}}=R0 \left(1-\sqrt{\ds I\over\ds I_{\rm{BEC}}}\right)^{1/2},
\ee
and for a thermal distribution, a more familiar expression, namely
\be\label{eq21}
R_{\rm{TF}}=R0 \left(1-\sqrt{\ds T\over\ds T_{\rm{BEC}}}\right)^{1/2}.
\ee
The precise determination of $I_{\rm{BEC}}$, $T_{\rm{BEC}}$ and $R0$ is however 
intimately related to the solution of Eq.(\ref{eq19}). 
}

\setcounter{equation}{0}
\bc{\section{Excitations of the Condensate}}\ec

The small excitations of the condensate (collective modes) are well studied with 
the RPA technique. Indeed, one may first set
\be\label{eq22}
\ba{rl}
\ord & = \ord0 +\delta\ord
\\
\nt  & = \nt0+\delta\nt
\\
\mt  & = \mt0+\delta\mt
,
\ea 
\ee
$\ord0$, $\nt0$ and $\mt0$ being the static solutions satisfying 
(\ref{eq9}), (\ref{eq12}) or (\ref{eq18}) according to the phase the system 
is in and $\delta\ord$, $\delta\nt$ and $\delta\mt$ are small deviations from 
local equilibrium. Then, one may expand the dynamical equations (\ref{eq7}) 
dropping terms up to second order, but keeping the kinetic terms since they 
are crucial for the excitation spectrum.

At zero temperature, we recall that the sole meaningful equation is the first
one in the eqs.(\ref{eq7}). Its expansion around the static solution
(\ref{eq9}) leads to the following equation:
\be\label{eq23}
i\frac{\hbar}{g}\delta\dot{\ord} = \left(
-{\ds\hbar^2\over\ds 2gm}\Delta + \ki +2\nc0\right)\delta\ord
+ \ord0^2\delta\ord^{*}. 
\ee
Upon setting $\ep=\hbar^2 k^2/2m$, we can solve for the modes 
$\delta\ord ({\bf r},t)=e^{-i\omega t} u({\bf r})+e^{i\omega t} v({\bf r})$ 
to obtain the dispersion relation for a uniform gas
\be\label{eq24}
\hbar\omega_k = \sqrt{\ep (\ep +2g\nc0)},
\ee
which is the Bogoliubov spectrum at $T=0$ \cite{ST}. For instance, at small 
momenta, one obtains the phonon spectrum $\omega_k = \sqrt{\frac{g\nc0}{m}} k$.

For finite temperatures and above the transition, we linearize the TDHFB eqs.(\ref{eq7}) 
without using the TF approximation  in order to keep the kinetic term $\epk=\ep/g$. The
procedure provides a ($4\times 4$) RPA system of the form: $i\frac{\hbar}{g}\dot{V}=MV$, 
where $V$ is the column vector $(\delta\ord,\delta\ord^*,\delta\mt,\delta\mt^*)$ and the 
RPA matrix is given by:
\be\label{eq26}
M=\pmatrix{M_1 & M_2 & M_{3}^{*} & 0\cr 
-M_{2}^{*} & -M_1 & 0 & -M_{3}\cr
M_4 & -M_5 & M_6 & 0 \cr
M_5^{*} & -M_4^{*} & 0 & -M_6 \cr
},
\ee
with
\be\label{eq26a}
\ba{rl}
M_1 & = \epk+\ki+2\nt0 \\
M_2 & = \mt0 -\ord0^2 \\
M_3 & = \ord0 \\
M_4 & = 2\ord0 (1+2\nt0-\nc0)-2\mt0\ord0^{*} \\
M_5 & = 2\ord0 (\mt0 +\ord0^2)\\
M_6 & = 4(\epk+\ki+2n)+1+2\nt0
.
\ea 
\ee

After some algebra, we obtain the expressions for the eigenfrequencies:
\be\label{eq27}
\omega_{\pm}={\ds g\over\ds\hbar}
\left\{
\demi \left(B\pm\sqrt{B^2-4E}\right)
\right\}^{1/2}
\ee
where
$$
B=4\nc0(1+2\nt0)+\left[4(M_1+2\nc0)+1+2n\right]\left[4(M_1+\nc0)+1+2n\right],
$$
and
$$
E=-8\nc0\left[1+2\nt0+3M_1+4\nc0\right]\left[M_1(1+2\nt0)+4(M_1+\nc0)(M_1+2\nc0)\right].
$$
The static solutions are computed from (\ref{eq18}) with $\ki$ replaced by $\epk+\ki$.

The spectrum (\ref{eq27}) clearly exhibits a departure from the Bogoliubov spectrum (\ref{eq24}). 
It is indeed temperature and position-dependent. The two modes $\omega_{\pm}$ reduce respectively 
to $\ep +V_{\rm{ext}}-\mu$ and $4(\ep +V_{\rm{ext}}-\mu)$ in the noninteracting case ($g=0$).
Therefore, one may say in the interacting case that $\omega_{-}$ and $\omega_{+}$ describe the 
coupled oscillations of the condensate fraction and the anomalous density. What remains is a 
comparison between these modes and existing data. But this requires a detailed knowledge of 
the static solution as shown by (\ref{eq27}). Indeed, the temperature dependence of the static 
solution complicates the expressions for the eigenfrequencies and one therefore cannot use the 
simplifications that appear in the $T=0$ case, as was performed in \cite{KT97,KT}.

The detailed study of the static solution as well as the eigenmodes is in progress and we will 
postpone the results to a future paper\cite{BC}.

\setcounter{equation}{0}
\bc{\section{Conclusions and Perspectives}}\ec

We have been concerned in this paper with a dynamical variational generalization of the
Gross-Pitaevskii equation, which takes into account the coupling of the condensate with 
the thermal cloud and with the anomalous density. We show that our derivation is 
consistent with all known approximations which go beyond the Gross-Pitaevskii approximation, 
namely, the Popov, the Beliaev and the HFB approximations.

The equations that we obtain are fully self-consistent, mainly because they not only introduce 
a dynamics of the thermal cloud and the anomalous density, but they allow also for a consistent 
feedback effect of these densities on the condensate fraction.

Instead of solving the full dynamical equations, we choose to 
focus first on the static situation, where theoretical works as well as experimental 
data do exist.

At zero temperature, we obtain familiar expressions for the chemical 
potential and the condensate radius. But for finite temperatures and above
the transition, the situation is much more complicated and requires a numerical
study.

The preliminary results show a good qualitative agreement 
with what is known; in particular, the anomalous density is always extremely small which is 
compatible with the Thomas-Fermi approximation.  

We then turn to the small amplitude motion and derive RPA-like equations which provide two coupled 
modes of oscillations; the well-known breathing modes of the condensate. A direct comparison with 
the results of \cite{KT97,KT} is unfortunately a little bit delicate since both the $T=0$ and the 
$V_{\rm{ext}}=0$ cases were considered there. We have shown in particular that, owing to a temperature 
dependence from the beginning, makes the $T=0$ limit a rather subtle question.

This paper is dedicated to the memory of Dominique Vautherin (Arthur), an active 
member of the Division de Physique Th\'eorique, Institut de Physique Nucl\'eaire (IPN), 
Orsay-France.

We are grateful to P. Schuck and M. V\'en\'eroni for fruitful discussions. We are particularly
indebted to C. Martin for her valuable comments and a careful reading of the manuscript.

\disregard{

\setcounter{equation}{0}
\renewcommand{\theequation}{A.\arabic{equation}}
{\Large{\bf Appendix}}

For the sake of completeness, we recall in this Appendix some useful
properties of bosonic gaussian operators of the form:
\be\label{a1}
{\cal T}={\cal N}\exp{(\la\tau\al)}\exp{(\demi\al\tau S\al)}.
\ee
In (\ref{a1}), $\cal N$ is a c-number, $\la (\x,t)$ a $2N-$component vector
and $S(\x,\y,t)$ is a $2N\times 2N$ symplectic matrix. The ($2N\times 2N$)
matrix $\tau$ is defined as
\be\label{a2}
\tau (\x,\y)=\pmatrix{0&\delta^D (\x-\y)\cr -\delta^D (\x-\y)&0\cr},
\ee
and $\al (\x)$ is the $2N-$component boson operator in the Schr\"odinger
picture
\be\label{a3}
\al_j (\x)=
{\ds 1\over\ds\sqrt{2}}
\left\{\ba{rl}
&\bphi_j (\x)+i\bpi_j (\x)\quad\quad\quad j=1,2\ldots N\\
&\bphi_{j-N}(\x)-i\bpi_{j-N}(\x)\quad j=N+1,\ldots 2N
\ea\right.
\ee
obeying the usual commutation relations
\be\label{a4}
[\al_i(\x) ,\, \al_j(\y)]=\tau_{ij}(\x,\y).
\ee
We have adopted in (\ref{a1}) compact notations for discrete sums and space
integrals. For instance
\be\label{a5}
\ba{rl}
&\la\tau\al =\sum_{i,j=1}^{2N}\int_{\x,\y}\la_i(\x,t)\tau_{ij}(\x,\y)\al_j
(\y),\\
&\al\tau S\al =\sum_{i,j,k=1}^{2N}\int_{\x,\y,\z}\al_i(\x)\tau_{ij}(\x,\y )
S_{jk}(\y,\z,t)\al_k(\z).
\ea
\ee

As mentioned in the text, it is more convenient to work with the set of three
parameters consisting of the ''partition function'' $\cal Z$, the vector
$\am$ and the contraction matrix $\r$. These are given by
\be\label{a6}
\left\{\ba{rl}
{\cal Z}\equiv & \rm{Tr}\,{\cal T}={\cal N}\exp{(\demi\la\tau\r\la)}
\sqrt{\det{\sigma\r}} ,\\
\am\equiv & \rm{Tr}\,{\cal T}\al/{\cal Z}={\ds 1\over\ds {1-\exp{(-S)}}}
\la ={\ds 1\over\ds 1-T}\la ,\\
\r \equiv & \rm{Tr}\,\big({\cal T}\tau\bar{\al}\ti{\bar{\al}}\big)/{\cal Z}
={\ds 1\over\ds{\exp{(S)}-1}}={\ds T\over\ds 1-T},
\ea\right.
\ee
where $T=e^{-S}$, $\bar{\al}=\al -\am$ and
\be\label{a7}
\sigma (\x,\y)=\pmatrix{0&\delta^D (\x-\y)\cr \delta^D (\x-\y)&0\cr}.
\ee

The product of two gaussian operators ${\cal T}={\cal T}_{1}{\cal T}_{2}$
of the form (\ref{a1}) with parameters ${\cal Z}_1$, $\am_1$, $\r_1$ and
${\cal Z}_2$, $\am_2$, $\r_2$, is also a gaussian operator of the form
(\ref{a1}) with parameters ${\cal Z}$, $\am$ and $\r$. These are given by
\be\label{a8}
\left\{\ba{rl}
{\cal Z} = & {\cal Z}_1 {\cal Z}_2 
\exp{\big\{\demi (\am_1-\am_2)\tau{\cal P}(\am_1-\am_2)\big\}}
\sqrt{\det{\sigma{\cal P}}} ,\\
\am = & \r_1{\cal P}\am_2+(\r_2+1){\cal P}\am_1 ,\\
\r = & \r_1 {\cal P}\r_2 ,
\ea\right.
\ee
with ${\cal P}=(1+\r_1+\r_2)^{-1}$. As an immediate consequence, a projector
on coherent states (that is a pure state density matrix):
${\cal T}^2={\cal Z}{\cal T}$ satisfies the property
\be\label{a9}
\r^2 +\r =0 .
\ee

The expressions of $\am$ and $\r$ become more intuitive in terms of the
usual averages defined as ($i,j=1,\ldots N$)
\be\label{a10}
\left\{\ba{rl}
\phi_i (\x,t) &=\langle\bphi_i(\x)\rangle ,\\
\pi_i  (\x,t) &=\langle\bpi_i (\x)\rangle ,
\ea\right.\quad
\left\{\ba{rl}
\g_{ij}(\x,\y,t) &=\langle\bar{\bphi}_i(\x)\bar{\bphi}_j(\y)\rangle ,\\
\ff_{ij}(\x,\y,t) &=\langle\bar{\bpi}_i(\x)\bar{\bpi}_j(\y)\rangle ,\\
\fipi_{ij}(\x,\y,t) &=\langle\bar{\bphi}_i(\x)\bar{\bpi}_j(\y)+
\bar{\bpi}_j(\y)\bar{\bphi}_i(\x)\rangle ,
\ea\right.
\ee
with $\bar{Q}=Q-\langle Q\rangle$. Indeed, with the help of the canonical
transformations (\ref{a3}), one gets easily
\be\label{a11}
\am ={\ds 1\over\ds\sqrt{2}}\left(\ba{c}
\phi +i\pi \\
\phi -i\pi\ea\right),
\ee
and
\be\label{a12}
\r =\demi\pmatrix{
\g+\ff-\frac{i}{2}(\fipi -\ti{\fipi})&
-(\g-\ff)-\frac{i}{2}(\fipi +\ti{\fipi})\cr
(\g-\ff)-\frac{i}{2}(\fipi +\ti{\fipi})&
-(\g+\ff)-\frac{i}{2}(\fipi -\ti{\fipi})\cr}.
\ee

}

\end{document}